\documentclass{sigchi}
\CopyrightYear{2018}
\setcopyright{acmlicensed}
\conferenceinfo{CHI 2018,}{April 21--26, 2018, Montreal, QC, Canada} \isbn{978-1-4503-5620-6/18/04}\acmPrice{\$15.00}
\doi{https://doi.org/10.1145/3173574.3174209}
\usepackage[T1]{fontenc}   
\usepackage{amsmath}
\usepackage{amssymb}
\usepackage{mathptmx}
\usepackage{balance}       
\usepackage{graphics}      
\usepackage{color}
\usepackage[pdflang={en-US},pdftex]{hyperref}
\usepackage{booktabs}
\usepackage{times}
\usepackage{caption}
\usepackage{microtype}        
\usepackage{ccicons}         
\usepackage{xspace}
\usepackage{enumitem}
\usepackage{float}
\usepackage{multirow}
\usepackage[subtle]{savetrees}
\usepackage{mathtools}
\usepackage{balance}

\DeclareRobustCommand{\etal}{\textit{et al.}\xspace}

\DeclareRobustCommand{\etal}{\textit{et al.}\xspace}
\DeclareRobustCommand{\surl}[1]{\emph{\urlstyle{same}\url{#1}}\xspace}

\DeclareRobustCommand{\etal}{\textit{et al.}\xspace}
 
\DeclareRobustCommand{\bfparhead}[1]{\noindent\textbf{#1}}   
\DeclareRobustCommand{\bp}{\emph{backward projection}\xspace}

\DeclareRobustCommand{\fp}{\emph{forward projection}\xspace}

\DeclareRobustCommand{\fm}{\emph{feasibility map}\xspace}
\DeclareRobustCommand{\pl}{\emph{prolines}\xspace}

\DeclareRobustCommand{\surl}[1]{\emph{\urlstyle{same}\url{#1}}\xspace}

\definecolor{orange}{rgb}{1, 0.49, 0.05}
\definecolor{gray}{rgb}{0.49, 0.49, 0.49}

\def\plaintitle{A Visual Interaction Framework for \\Dimensionality Reduction Based Data Exploration}

\def\emptyauthor{}
\def\plainkeywords{
  Dimensionality reduction; 
  interaction; 
  bidirectional binding; 
  visual embedding; 
  forward projection; 
  backward projection; 
  PCA; 
  autoencoder; 
  deep learning; 
  proline; 
  feasibility map; 
  exploratory data analysis;
  what-if analysis; 
  Praxis.}

\makeatletter
\def\url@leostyle{%
  \@ifundefined{selectfont}{
    \def\UrlFont{\sf}
  }{
    \def\UrlFont{\small\bf\ttfamily}
  }}
\makeatother
\def\pprw{8.5in}
\def\pprh{11in}

\definecolor{linkColor}{RGB}{6,125,233}
\hypersetup{%
  pdftitle={\plaintitle},
  pdfauthor={\emptyauthor},
  pdfkeywords={\plainkeywords},
  pdfdisplaydoctitle=true, %
  bookmarksnumbered,
  pdfstartview={FitH},
  colorlinks,
  citecolor=black,
  filecolor=black,
  linkcolor=black,
  urlcolor=linkColor,
  breaklinks=true,
  hypertexnames=false
}

\begin{document}
\title{\plaintitle}

\numberofauthors{2}
\author{%
\alignauthor{Marco Cavallo\\
\affaddr{IBM Research}\\
\email{mcavall@us.ibm.research}}\\
\alignauthor{\c{C}a\u{g}atay Demiralp\\
\affaddr{IBM Research}\\
\email{cagatay.demiralp@us.ibm.research}}\\
}

\maketitle
\begin{abstract}
  Dimensionality reduction is a common method for analyzing and visualizing
high-dimensional data. However, reasoning dynamically about
the results of a dimensionality reduction is difficult. Dimensionality-reduction
algorithms use complex optimizations to reduce the number of dimensions of a
dataset, but these new dimensions often lack a clear relation to the initial data
dimensions, thus making them difficult to interpret.

Here we propose a visual interaction framework to improve
dimensionality-reduction based exploratory data analysis. We introduce two
interaction techniques, forward projection and backward projection, for
dynamically reasoning about dimensionally reduced data. We also contribute two
visualization techniques, prolines and feasibility maps, to facilitate the
effective use of the proposed interactions.

We apply our framework to PCA and autoencoder-based dimensionality reductions.
Through data-exploration examples, we demonstrate how our visual interactions
can improve the use of dimensionality reduction in exploratory data analysis.

\end{abstract}


\section{Introduction}

Dimensionality reduction (DR) is widely used for exploratory data analysis
(EDA)\cite{Tukey_1966} of high-dimensional datasets.  DR algorithms
automatically reduce the number of dimensions in data while maximally
preserving structures, typically quantified as similarities, correlations or
distances among data points~\cite{Maaten_2009}. This makes visualization of the
data possible using conventional spatial techniques. For instance, analysts
generally use scatter plots to visualize the data after reducing the number of
dimensions to two, encoding the reduced dimensions in a two-dimensional
position. 

\begin{figure}[h]
  \includegraphics[width=\linewidth]{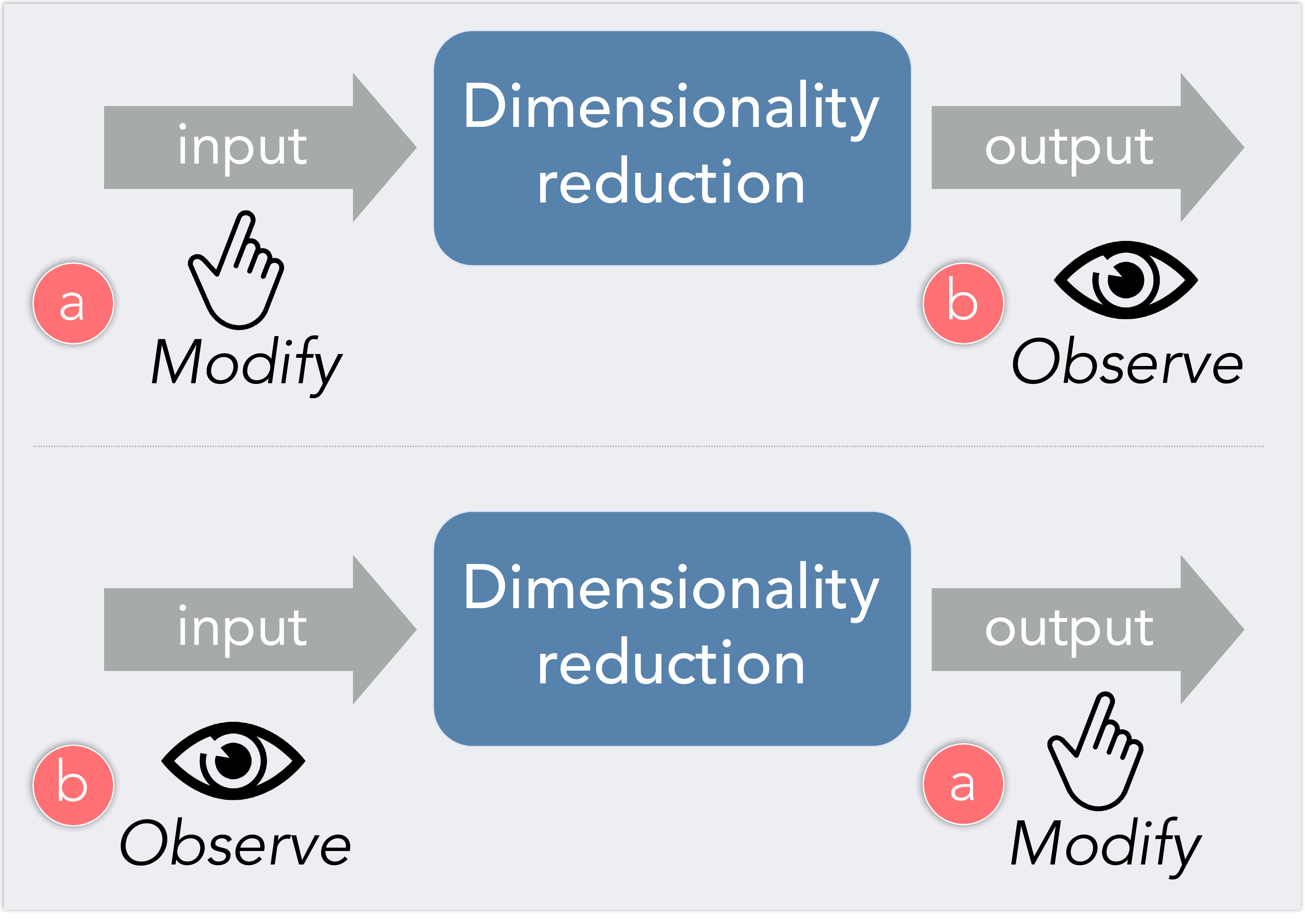}
  \caption{Better interaction with DR models to improve DR-based 
    exploratory analysis? Two complementary affordances can help: (Top) Modify
    the input to the  model first and then see how the model output changes.
    (Bottom) Modify the model output first and then see how the input needs to
    change (be synthesized) so that the DR model generates the user-induced
    output. These two affordances can improve DR-based exploratory analysis 
    by concurrently facilitating model understanding and what-if analysis.
    \label{fig:howto}} 
\end{figure}

\bfparhead{DR Challenges:} DR methods are driven by complex numerical
optimizations, which makes dynamic reasoning about DR results difficult.
Dimensions derived by these methods generally lack clear, easy-to-interpret
mappings to the original data dimensions.  Data analysts with limited
experience in DR have in difficulty in interpreting the meaning of the
projection axes and the position of scatter plot nodes~\cite{Brehmer_2014,Sedlmair_2013}: 
`What do the axes mean?'' is frequently asked by users looking
at scatter plots whose points (nodes) correspond to dimensionally reduced data.
As a result, analysts treat DR methods as black boxes and often rely on
off-the-shelf routines in their toolset for computation and visualization.
Indeed, most scatter-plot visualizations of dimensionally reduced data are
viewed as static images. One reason is that tools for computing and plotting
these visualizations, such as Python, R, and Matlab, have few  interactive
exploration functionalities. Another reason is that few interaction and
visualization techniques go beyond brushing-and-linking or cluster-based
coloring to allow dynamic reasoning with DR visualizations. 

However, the ability not only to probe the results of a DR model 
but also actively to tinker with them is important for a model-based EDA,
as experimentation is essential to data exploration~\cite{Tukey_1966}. 
Enabling an analyst to run various input and output scenarios and see 
how the underlying DR model---coupled with data---responds can facilitate 
model understanding and is a prerequisite for what-if analysis.   

\bfparhead{Improving DR-Based Exploratory Analysis:} In response, we propose a
visual interaction framework to improve DR-based data exploration. To this
end, we introduce two interaction techniques, \fp and \bp, to help analysts
dynamically explore and reason about scatter plot representations of
dimensionally reduced data.  We also contribute two visualization techniques,
\pl and \fm, to facilitate the effective use of the proposed interactions.

The underlying idea (Figure~\ref{fig:howto}) of our framework is that the
ability to induce \textit{change} and observe the \textit{effects} of that
change is essential for reasoning  with DRs or any black-box models (more on
this in Discussion). Forward projection enables an analyst to interactively
change data attributes input to a DR routine and observe
the effects in the output. Backward projection complements forward projection
by letting the analyst make hypothetical changes to the output, the
attributes of new dimensions, and observe which changes in the input attribute
values would produce the hypothesized changes in the output.  These affordances
are useful for running what-if scenarios as well as  understanding the
underlying DR process.  

\bfparhead{Contributions:} Our high-level contribution is 1) a new framework
that aims to enable  users to dynamically change the input and output of DRs
and observe the effects of these changes. The design of the visual interactions
that operationalize the underlying purpose of the framework also has novel
attributes. These contributions include 2) the forward projection interaction
using out-of-sample (OOS) extension, 3) the proline visualization along with
its visual and interactive affordances, 4) the backward projection interaction
with interactive user constraints, and 5) the feasibility map visualization.  

Any DR algorithm with fast OOS extension (extrapolation) and inversion methods
can be plugged in our framework.  We apply the framework to PCA (principal
component analysis) and autoencoder-based dimensionality reductions and
demonstrate how it improves DR-based exploratory analysis.     

Next we discuss related work and then introduce  our
framework interactions. We then present applications to PCA and autoencoder and
give exploratory analysis examples, and then  elaborate on the scalability and
accuracy of our methods. Then we discuss how the current model extends to
black-box models at large, such as deep learning models, and how changes in model
development practices can help improve explorability and interpretability of
black-box models. We conclude by summarizing  our contributions and reflecting
on the importance of EDA tools that support interactive experimentation.

\section{Related Work}

Our work builds on prior research in direct manipulation and auxiliary visual 
encoding in scatter plots of dimensionality reductions (DRs).   

\subsection{Direct Manipulation in DR} Direct manipulation has a long history
in human-computer
interaction~\cite{borning1981programming,kay1977personal,sutherland1964sketchpad}
and visualization research (e.g.~\cite{shneiderman1982direct}). Direct
manipulation techniques aim to improve user engagement by minimizing the
\textit{perceived} distance between the interaction source and the target
object~\cite{hutchins1985direct}.

Developing direct manipulation interactions to guide DR formation and modify
the underlying data is a focus of prior research~\cite{Buja_2008,Endert_2012,
Gleicher_2013,Jeong_2009,Johansson_2009,Williams_2004}.  For example,
X/GGvis~\cite{Buja_2008} supports changing the weights of dissimilarities input
to the MDS stress function along with the coordinates of the embedded  points
in order to guide the projection process. Similarly, iPCA~\cite{Jeong_2009}
enables users to interactively modify the weights of data dimensions in
computing projections. Endert \etal~\cite{Endert_2011} apply similar ideas to
additional dimensionality-reduction methods while incorporating user feedback
through spatial interactions in which users can express their intent by
dragging points in the plane.

Earlier work also uses direct manipulation to modify data through DR
visualizations in order to support, e.g., exploratory analysis
~\cite{Jeong_2009}, multivariate network manipulation~\cite{viau2010flowvizmenu},
exploration of trajectory clusters~\cite{Schreck_2009}, movement trace
analysis~\cite{crnovrsanin2009proximity}, and feature transformation
~\cite{mamani2013user}.   Our work here aims to facilitate DR-based exploratory
analysis. Akin to forward projection and unconstrained backward projection
techniques, iPCA~\cite{Jeong_2009} enables interactive forward and backward
projections for PCA-based DRs.  However, iPCA recomputes full PCAs for each
forward and backward projection, and these can suffer from jitter and
scalability issues.  Using out-of-sample
extrapolation~\cite{Bengio_2004,Maaten_2009}, our forward projection avoids
re-running dimensionality reduction algorithms. Unlike iPCA, we also enable
users to interactively define constraints on feature values and perform
constrained backward projection. 

We refer readers to a recent survey~\cite{sacha2017visual} for an exhaustive
discussion of prior work  on visual interaction with dimensionality reduction. 

\subsection{Visualization in DR Scatter Plots} Prior work incorporates various
visualizations in planar scatter plots of DRs in order to improve the user
experience~\cite{Aupetit_2007,Chuang_2012,clustrophile:idea16,gabriel1971biplot,
Jeong_2009,Lespinats_2010,Stahnke_2016}. 
Since low-dimensional projections are generally lossy  representations of the
high-dimensional data relations, it is useful to convey both overall and
per-point dimensionality-reduction errors to users when desired.  Researchers
visualized errors in DR scatter plots using Voronoi
diagrams~\cite{Aupetit_2007,Lespinats_2010} and corrected (undistorted) the
errors by adjusting the projection layout with respect to the examined
point~\cite{Chuang_2012,Stahnke_2016}.

Biplot was introduced~\cite{gabriel1971biplot} to visualize the magnitude and
sign of a data attribute's contribution to the first two or three principal
components as line vectors in PCA. Biplots are  computed using singular-value
decomposition, regardless of the actual DR used, assuming the underlying DR is
linear and the data matrices needed to compute the decomposition are
accessible. 

Closest to our prolines are the \textit{enhanced biplots} introduced by
Coimbra \etal~\cite{coimbra2016explain}. Enhanced biplots aim to extend biplots
to nonlinear DRs and assume only access to the projection function of a DR,
thus sharing similar assumptions and generalization properties with prolines.
Similarly  to a proline construction, each axis of an enhanced biplot is
constructed by connecting the projections of points sampled on the range of the
corresponding data attribute. However, prolines differ from enhanced biplots in
a few aspects.  Both enhanced and classical biplots visualize how projections
(reduced dimensions) change on average with changing attribute values, whereas
Prolines are computed for each data and attribute and visualize projection
changes locally for each data point. In this sense, prolines complement
enhanced biplots by constructing local axes of projection change with respect
to data attributes. To construct an attribute axis, enhanced biplots use values
regularly sampled on the attribute's range with a sampling rate uniform across
axes, while keeping  the remaining attributes constant at their average values.
On the other hand, we modulate the sampling rate with attribute variances and
decorate prolines with marks to communicate distributional characteristics of
the underlying data point. More crucially, prolines  differ from this earlier
work in being interactive visual signifiers that dynamically facilitate user
interactions. 

Stahnke \etal~\cite{Stahnke_2016} use a grayscale map to visualize how a single
attribute value changes between data points in DR scatter plots.  We introduce
the feasibility map, a grayscale map, to visualize the feasible regions in the
constrained backward projection interaction.

We have presented  our work at different stages of its development at two
workshops.  We introduced an initial version as part of Clustrophile, an
exploratory visual clustering analysis tool~\cite{clustrophile:idea16}. We then
presented our revised visual interactions integrated with Praxis, an
interactive DR-based exploratory analysis tool, in a dedicated
draft~\cite{Cavallo:2017:IDEA}. Here we give a unified treatment of our work by
formalizing it under a framework. The current work also demonstrates the use of
our visual interactions through several new data-exploration examples and
provides a new discussion that relates the applicability of our framework to
black-box models, particularly deep learning models, at large.

\section{Visual Interactions} 
We now discuss the interactions and the related visualizations in our framework.  
\begin{figure}
 \centering
 \includegraphics[width=1\linewidth]{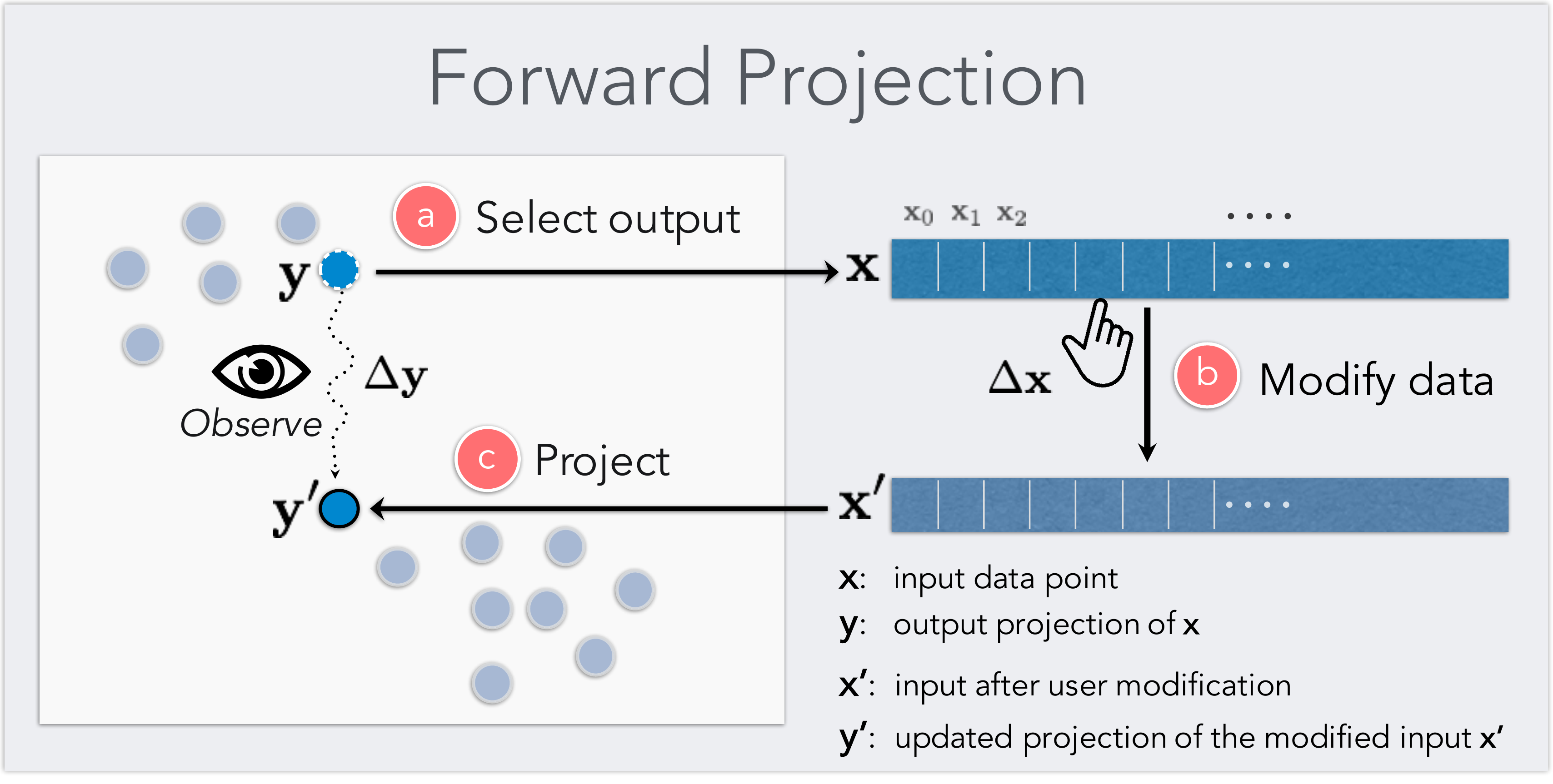}
 \caption{\normalfont   Forward projection enables users to: (a) select any data 
   point instance $\mathbf{x}$ that is input to a DR, (b) interactively change 
   its high-dimensional feature values, and (c) observe the change $\Delta \mathbf{y}$ 
   in the point's two-dimensional projection.\label{fig:fp}}
\end{figure}
\subsection{Forward Projection} 
Forward projection enables users to interactively change the feature values  of a data input
$\mathbf{x}$ and observe how these hypothesized changes in data modify the
current projected location $\mathbf{y}$ (Figure~\ref{fig:fp}). 

We compute forward projections using out-of-sample (OOS) extension (or
extrapolation)~\cite{Maaten_2009}. OOS extension is the projection of a new
data point into an existing dimensionality reduction (DR) using only 
the properties of the already computed DR.  It is thus 
conceptually equivalent to testing a trained machine-learning model with  data
that was not part of the training set. Most common DR methods have OOS
extension algorithms with desirable accuracy properties~\cite{Bengio_2004}.

We propose using OOS extension as opposed to re-running the DR for two basic
reasons.  The first is scalability: OOS computation is  generally much
faster than re-running the dimensionality reduction, and speed is critical  in
sustaining the interactive experience.  The second is preserving the constancy of
scatter plot representations~\cite{bederson1999does}. For example, re-running
(training) a dimensionality-reduction algorithm with a new data sample added
can significantly alter the two-dimensional scatter plot of the dimensionally
reduced data, even though all the original inter-datapoint similarities may
remain unchanged. With OOS, forward projection animations change the position
of only the point attributes that  the user interactively modifies.   

\subsection{Prolines: Visualizing Forward Projections}
Forward projection provides a scalable interaction to change the attributes of
a data instance and see how the dimensionality reduction changes. We introduce
prolines  to let users see  in advance  what forward projection paths look like
for each data point and feature. Through prolines, an analyst can see what
directly start exploring data without considering forward projections
exhaustively.   

\begin{figure}[t]
  {\centering
\includegraphics[width=\linewidth]{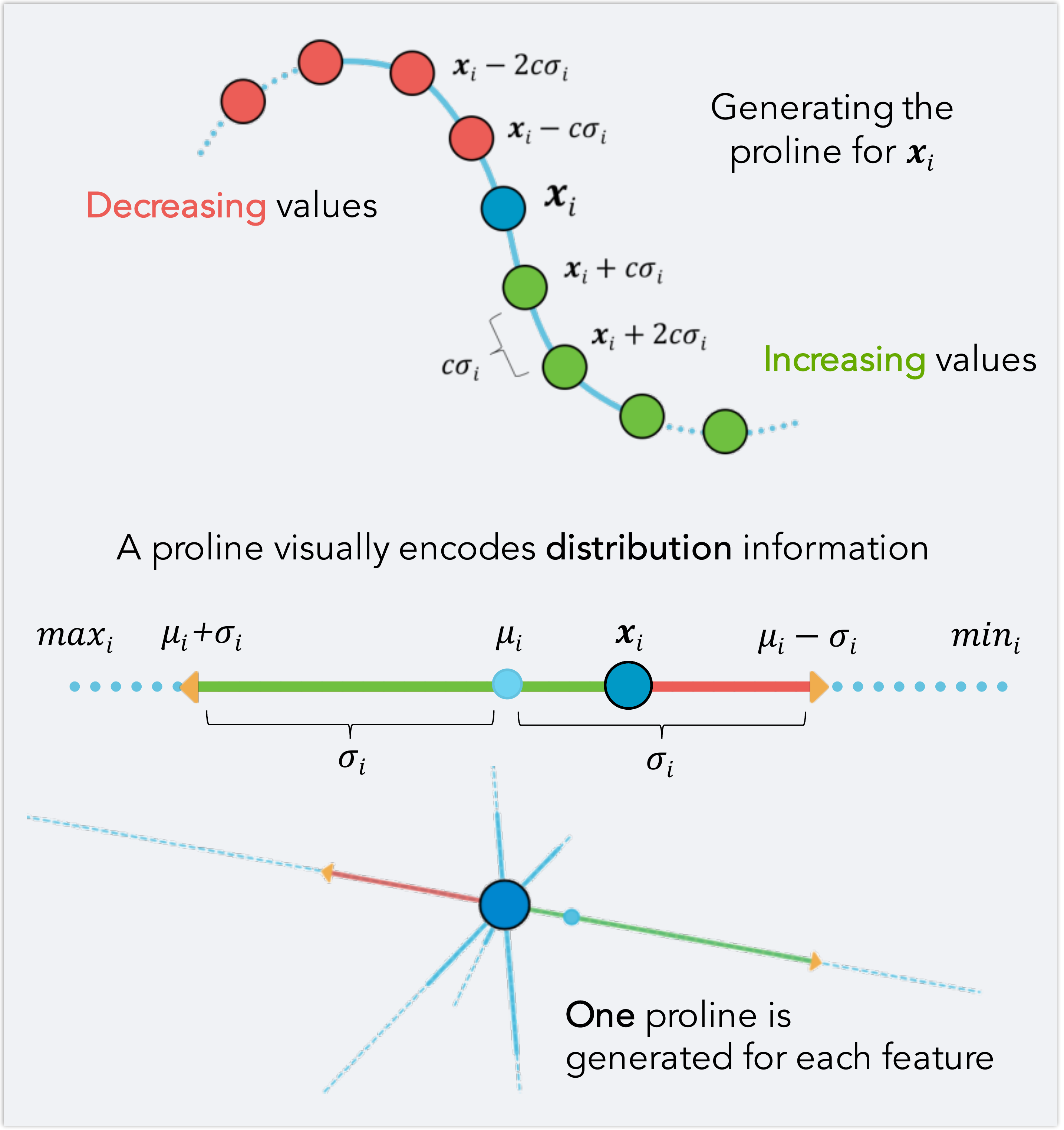}
\caption{ \normalfont Proline construction. For a given dimension (feature) ${x}_i$ of
  a point $\mathbf{x}$ in a dataset, we construct a proline by
  connecting the forward projections of data points regularly sampled from a
  range of $\mathbf{x}$ values, where all features are fixed but ${x}_i$
  varies. A proline also encodes the forward projections for the ${x}_i$ values
  in $\left[\mu_i-\sigma_i,\;\mu_i+\sigma_i\right]$ with thick green and red
  line segments, providing a basic directional statistical context. $\mu_i$ is
  the mean of the $i$th dimension in the dataset, the green segment represents
  forward projections for ${x}_i$ values in $\left[{x}_i,
  \mu_i+\sigma_i\right]$, and the red segment represents ${x}_i$ values in
  $\left[\mu_i-\sigma_i,\; {x}_i\right]$.\label{fig:pl}}
}
\end{figure}

Prolines visualize forward projection paths using regularly sampled values for
each feature and data point (Figures~\ref{fig:pl}).  Let $\mathbf{x}_i$ be the
value of the $i$th feature for the data point $\mathbf{x}$. We first compute
the mean $\mu_i$, standard deviation $\sigma_i$, minimum $\min_i$ and maximum
$\max_i$ values for the feature in the dataset  and devise a range $I =
\left[\min_i,\;\max_i\right]$. We then iterate over the range with step size
$c\sigma_i$, compute the forward projections as discussed above, and then
connect them as a path. 

In addition to providing an advance snapshot of forward projections, a proline
also conveys the relationship between the feature distribution and  the
projection space.  To that end, we display along  each proline a small
light-blue circle indicating the position that the data point would  assume if
it had a feature value corresponding to the mean of its distribution;
similarly, we display two small arrows indicating a variation of one standard
deviation ($\sigma_i$) from the mean ($\mu_i$). The segment identified by the
range $\left[\mu_i - \sigma_i, \mathbf{x}_i + \sigma_i\right]$ is highlighted
and further divided into two segments. The green segment shows the positions
that the data point would assume if its feature value increased; the red one
indicates a decreasing value. This enables users to infer the relationship
between the feature space and the direction of change in the projection space. 

\begin{figure}
\includegraphics[width=1\linewidth]{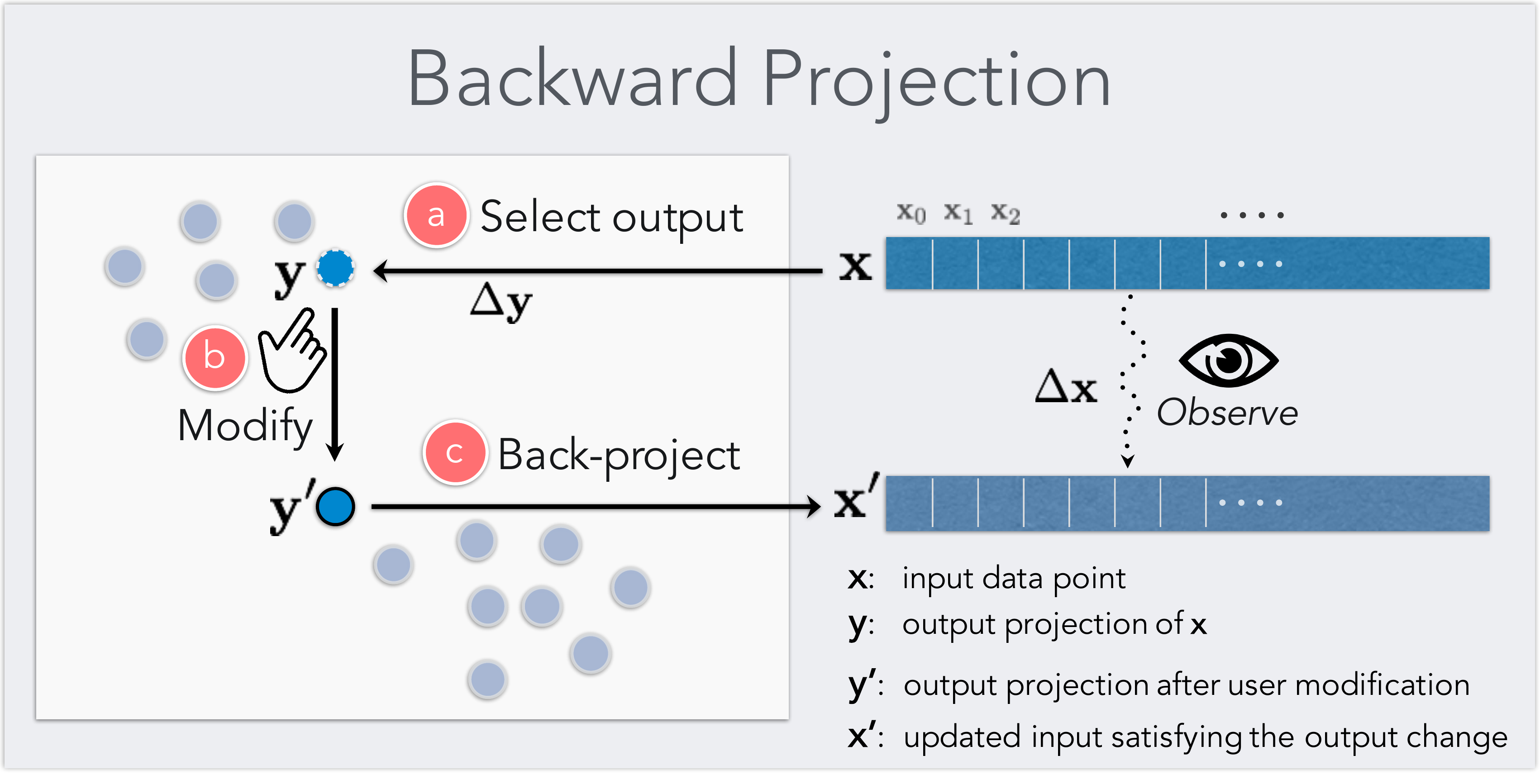}
\caption{Backward projection enables users to: (a) select any node
    corresponding to the two-dimensional projection of a data point
    $\mathbf{x}$, (b) move the node arbitrarily in the plane, and (c) observe
    the change $\Delta \mathbf{x}$ in the point's high-dimensional feature
    values.\label{fig:bp}}
\end{figure}  

\subsection{Backward Projection}

Backward projection complements the forward projection interaction by enabling
a user to interactively change output attributes and observe how the input
attributes change as the DR routine produces the user-induced output.
Consider the following scenario: a user looks at a projection and, seeing a
cluster of points and a single point projected far from this group, asks what
changes in the feature values of the outlier point would bring it near
the cluster. Now the user can play with  different dimensions using forward
projection interactions to move the current projection of the outlier point
near the cluster. It would be more natural, however, to move the point directly
and observe the change. 

Back or backward projection maps a low-dimensional data point back into the
original high-dimensional data space. For linear DRs, back projection is
typically done by applying the inverse of the learned linear DR mapping. For
nonlinear DRs, earlier research proposed DR-specific backward-projection
techniques. For example, iLAMP~\cite{dos2012ilamp} introduces a back-projection
method for LAMP~\cite{joia2011lamp} using local neighborhoods and demonstrates
its viability over synthetic datasets~\cite{dos2012ilamp}. Researchers also
investigated general backward-projection methods  based on radial basis
functions~\cite{amorim2015facing,monnig2014inverting}, treating backward
projection as an interpolation problem.
Autoencoders~\cite{hinton2006reducing}, neural-network-based DR models, are a
promising approach to computing backward projections. An autoencoder model with
multiple hidden layers can learn a nonlinear dimensionality-reduction function
(encoding) as well as the corresponding backward projection (decoding) as part
of the DR process.  

We propose both \textit{constrained} and \textit{unconstrained} backward
projection interactions.  Constrained backward projection enhances what-if
analysis by letting analysts semantically regulate the mapping into
unprojected high-dimensional data space. For example, we don't expect an Age
field to be negative or greater than 200, even though such a value can be a more
optimal solution in an unconstrained backward projection scenario. DRs are
many-to-one functions (more on this in Supplementary Material) in general, and
hence inverting them is an underdetermined problem that benefits from
regularization. Therefore, in addition to augmenting what-if analysis, the
ability to define constraints over a back projection can also ease the
computational burden by restricting the search space for a feasible solution.

It is important to note that, since more than one data point in the
multidimensional space can project to the same position, forward and backward
projections may not always correspond. For this reason, we add to our prolines
visualization a set of \textit{projection marks} (Figure~\ref{fig:oecd}b)
indicating the current value for each feature while the user performs backward
projection. At the same time, dragging a data point highlights the green or red
segment of each proline based on the increase or decrease of each feature,
showing which dimensions are correlated.  By combining forward
projection paths and backward projection, the user can infer how fast each value
is changing in relation to its feature distribution.


\begin{figure}[t]
\includegraphics[width=\linewidth]{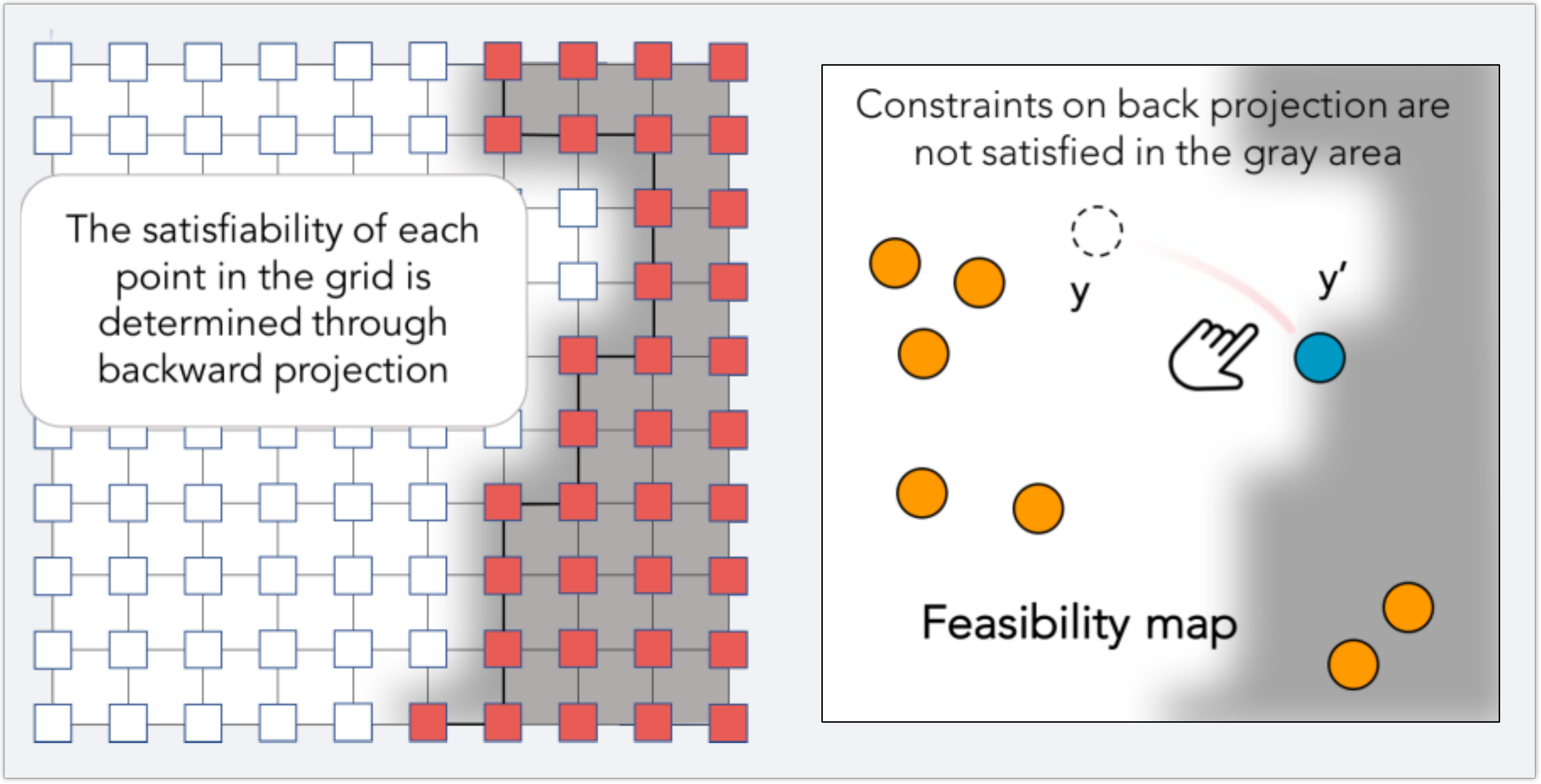}
\caption{\normalfont Feasibility map. The feasibility map is constructed by
  sampling the projection plane through constrained backward projection and
  then verifying the existence of each solution (left). The darker area of the
  map, computed through interpolation, corresponds to the positions of the
  plane that would break the constraints for the specified data point (right).
  \label{fig:fm}}
\end{figure}

\subsection{Feasibility Map: Visualizing Constraint Satisfaction}  
We propose the feasibility map visualization as a way quickly to see the feasible space
determined by  a given set of constraints.  Instead of manually checking if a
position in the projection plane satisfies the desired range of values
(considering both equality and inequality constraints), one would like  to know
in advance which regions of the plane correspond to admissible solutions. In
this sense, a  feasibility map is a conceptual generalization of prolines to the
constrained backward projection interaction.

To generate a feasibility map, we sample the projection plane on a regular grid
and evaluate the feasibility at each grid point based on the constraints
imposed by the user, obtaining a binary mask over the projection plane.  We
render this binary mask over the projection as an interpolated grayscale
heatmap in which darker areas indicate infeasible planar regions
(Figure~\ref{fig:fm}). With accuracy determined by the grid resolution, the
user can see which areas a data point can assume in the projection plane
without breaking the constraints.  In backward projection, if a data point is
dragged to a position that does not satisfy a constraint, its color and the
color of its corresponding projection marks turn to black. If the user drops
the data point in an infeasible position, the point is automatically moved
through animation back to the last feasible position to which it was dragged.

\begin{figure*}[tbh] \centering
  \includegraphics[width=0.9\linewidth]{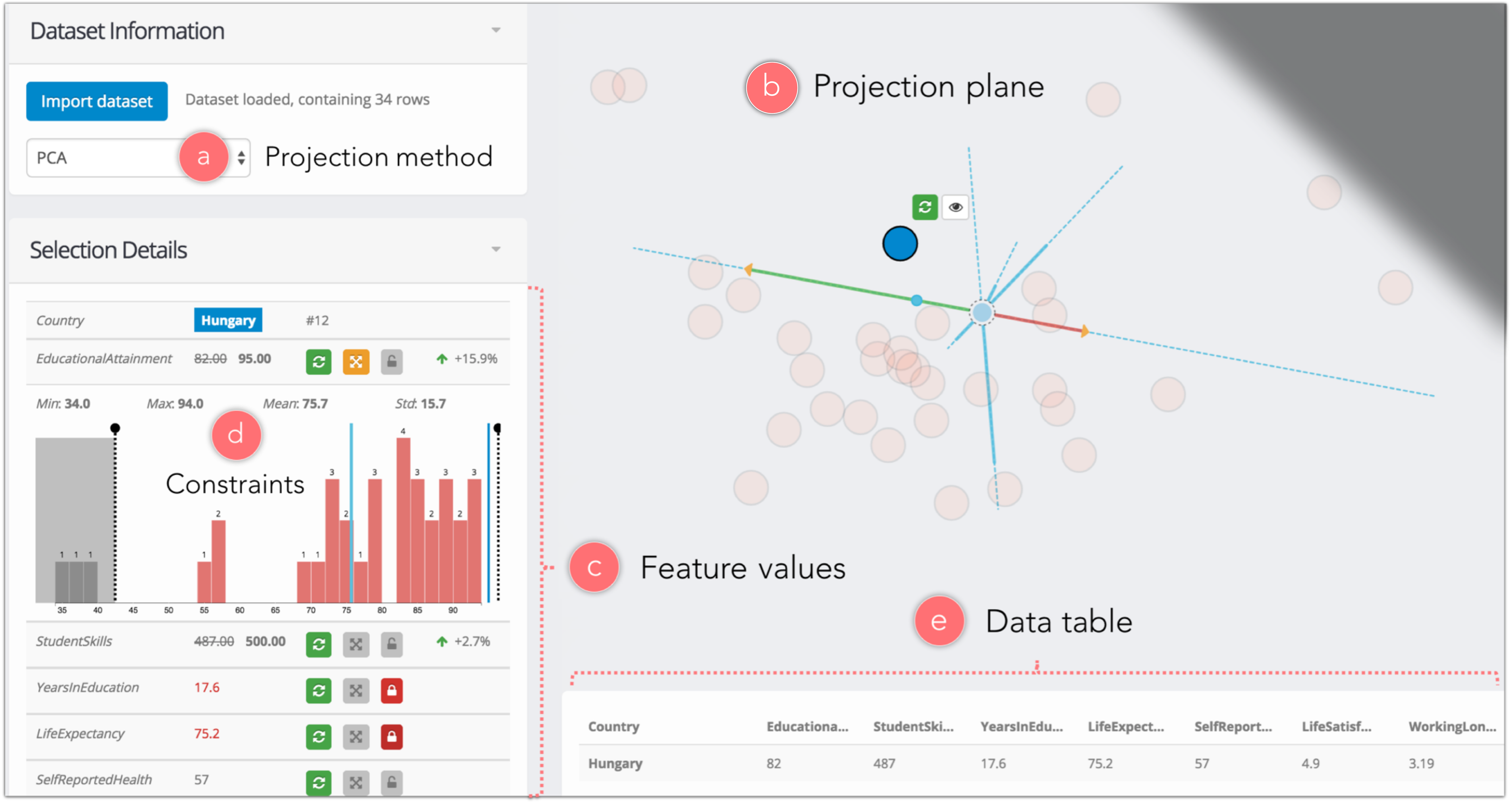}
  \caption{\normalfont Praxis interface. Praxis is a novel interactive tool for
    DR-based exploratory data analysis that integrates our visual interactions.  
    After importing a dataset and choosing a projection method
    (a), a scatter plot is displayed using the two reduced dimensions (b). When
    a point is selected, its feature values can be seen and modified from a
    table panel (c) that also allows entering constraints for each feature by
    double-clicking on a specific row of the table (d). A data table listing
    all rows in the dataset is also included (e). \label{fig:praxis}}
  \end{figure*}

\section{Applications}

We apply our framework to PCA (principal component analysis) and
autoencoder-based dimensionality reductions (DRs) and demonstrate how it
improves DR-based exploratory analysis. We choose PCA and autoencoder because
they respectively cover linear and nonlinear DR cases and have effective
extension and inversion methods. PCA, among the most frequently used DR
methods, is effective for rapid initial exploratory analysis and requires no
parameter tuning.  Note that the framework can be applied to any DR algorithm
with fast inversion and out-of-sample (OOS) extension methods. As their fast
inversion and extension methods become available, the framework can easily
applied to other popular DR methods such as t-SNE~\cite{maaten2008visualizing}.

In what follows we first briefly introduce Praxis, a new tool for interactive
DR-based data analysis  that integrates our framework interactions, and then
discuss applications through examples of exploration of tabular and image
datasets.

\bfparhead{Praxis:}\label{sec:praxis} To demonstrate the use of 
our interaction and visualization techniques, we
integrate them in Praxis, an interactive tool for DR-based exploratory
analysis. Although design and implementation details of Praxis are out of
the scope of this paper, we give a brief
description to help the reader follow the rest of the paper. 

Praxis' user interface has four basic elements: \textit{Data Import}, 
\textit{Selection Details}, \textit{Projection}, and \textit{Data Table}.  
Through the \textit{Data Import} panel (Figure~\ref{fig:praxis}a), users can
import a dataset in CSV format, set the projection method (e.g., PCA)
using a drop-down selection menu, and view the resulting dimensionality
reduction as a scatter plot in the \textit{Projection} plane  
(Figure~\ref{fig:praxis}b). As customary, Praxis uses the first two reduced
dimensions (e.g., the first two principal components for PCA) as axes.   

The results of forward and backward projection, along with the two visualizations
prolines and feasibility map, are displayed in the \textit{Projection} plane.
The id (name) of a data point is shown on mouse hover, while clicking performs
selection, showing its feature values in the \textit{Selection Details} panel,
a dedicated sidebar(Figure~\ref{fig:praxis}c). The \textit{Selection Details}
panel is used to perform forward projections (clicking on a dimension makes
its value modifiable) and to inspect changes in feature values when
backward projection is used. The three buttons next to the feature column
in this panel respectively 1) reset the feature to its initial value, 2) toggle
the inequality constraints on the feature, and 3) lock the feature value to the
current---modified---value (i.e., toggles the equality constraint).

Double-clicking the row associated with a feature displays a histogram
representing its distribution below the selected row, showing some basic
statistics (Figure~\ref{fig:praxis}d). The current value of the feature is
represented by a blue line and a cyan line indicates the distribution mean.
Bins of the histogram are colored similarly to prolines: green for increasing
values and red for decreasing values with respect to the original feature value.
Users can set constraints on the feature  through direct manipulation in the
histogram visualization. Dragging one of the two black handles lets the user
set or unset lower and upper bounds for a feature distribution, 
thus defining a set of constraints for a specific data point.

Finally, selecting a data point in the \textit{Projection} plane displays two
buttons that respectively enable 1) resetting its feature values (and position)
to their original value and 2) showing a tooltip on top of its $k$ currently
nearest neighbors, in order to facilitate reasoning about similarity with
other data samples; this is particularly useful when performing back
projection.

\begin{figure*}[tbh] \centering
  \includegraphics[width=1\linewidth]{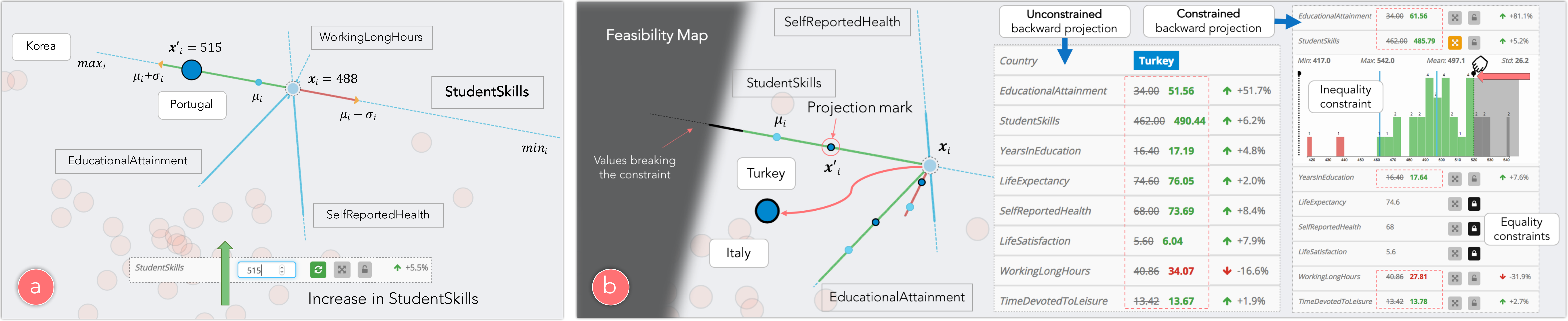} \caption{\normalfont
    PCA-based usage examples with the OECD Better Life dataset.
  \textbf{Left} (a): forward projection with prolines. \textsc{StudentSkills}
  is revealed as key feature differentiating Portugal from Korea. Observe that
  a value of 515 for \textsc{StudentSkills} would be reasonable with respect to
  the feature distribution ($\mu_i < 515 < \mu_i+\sigma_i$), but is not enough
  to make Portugal close to Korea in the projection plane. By visually
  comparing the lengths (variability) of different proline paths, the user can
  easily recognize which dimensions contribute most to determining the position
  of points in the dimensionally reduced space.
  \textbf{Right} (b): backward projection usage. Curious about the projection
  difference between Turkey and Italy, similar countries in some respects, the
  user moves the node associated with Turkey (blue circle) towards Italy.
  Turkey's feature values are automatically updated to satisfy the new
  projected position as the node is moved. The first table shows how its values
  would update with unconstrained backward projection, while the second table
  shows the result of constrained backward projection. \label{fig:oecd}}
\end{figure*}

We integrate PCA and autoencoder as dimensionality-reduction methods in Praxis.
Next we discuss the first application of our framework, PCA.  

\subsection{Application 1: PCA}

Principal component analysis (PCA) is  one of the most frequently used linear
dimensionality-reduction techniques. PCA computes (learns) a linear orthogonal
transformation of the empirically centered data into a new coordinate frame in
which the axes represent maximal variability. The process is identical to
fitting a high-dimensional ellipsoid to the data. The orthogonal axes of the
new coordinate frame, which are also the principal axes, are called principal
components.  To reduce the number of dimensions to two, for example, we project
the centered data matrix, rows of which correspond to data samples and columns
to features (dimensions), onto the first two principal components. Details of
PCA along with its many formulations and interpretations can be found in
standard textbooks on machine learning or data mining (e.g.,
\cite{Bishop_2006,Hastie_2005}). 

To compute the forward projection change $\Delta\mathbf{y}$ for PCA, we project
the data change vector $\Delta\mathbf{x}$ onto the first two principal
components: $\Delta\mathbf{y} = \Delta \mathbf{x}   \;\mathbf{E}$, where
$\Delta \mathbf{x}$ and $\Delta \mathbf{y}$ are row vectors, $\mathbf{E} =
\begin{bmatrix}\mathbf{e_0} & \mathbf{e_1} \end{bmatrix}$, and $\mathbf{e_0}$
and $\mathbf{e_1}$ are the first two principal components as column vectors.
The formulation of backward projection is the same as for forward projection:
$\Delta \mathbf{y} = \Delta \mathbf{x}\;\mathbf{E}$. In this case, however,
$\Delta \mathbf{x}$ is unknown and we need to solve the equation.  

In the case of unconstrained backward projection, we find $\Delta \mathbf{x}$
by solving a regularized least-squares optimization problem: 
\begin{equation*}
  \begin{aligned}
    & \underset{\Delta\mathbf{x}}{\text{minimize}}
    & & \|{\Delta\mathbf{x}}\|^2 \\
    & \text{subject to}
    & & \Delta\mathbf{x}\;\mathbf{E} = \Delta\mathbf{y}
  \end{aligned} 
\end{equation*}
We find a least-norm solution ${{\Delta \mathbf{x}}^{*}}$  by multiplying $\Delta
\mathbf{y}$ with the pseudoinverse of $\mathbf{E}$~\cite{boyd2004convex}. 
This is equivalent to setting ${{\Delta \mathbf{x}}^{*}}=\Delta\mathbf{y}\;\mathbf{E}^T$ 
as the pseudoinverse of a real-valued orthonormal matrix is equal to the transpose of
the matrix.    

For constrained backward projection, we find ${\Delta \mathbf{x}}^{*}$
by solving the following quadratic optimization problem:
\begin{equation*}
  \begin{aligned}
    & \underset{ \Delta\mathbf{x}}{\text{minimize}} & & \|{\Delta\mathbf{x}}\;\mathbf{E} - \Delta\mathbf{y}\|^2 \\
    & \text{subject to} & & \mathbf{C}\Delta\mathbf{x}  = \mathbf{d}\\
    & & & \mathbf{lb} \leq \Delta\mathbf{x} \leq \mathbf{ub} 
  \end{aligned}
\end{equation*}
Here $\mathbf{C}$ is the design matrix of equality constraints, $\mathbf{d}$ is the
constant vector of equalities, and $\mathbf{lb}$ and $\mathbf{ub}$ are the vectors
of lower and upper boundary constraints.

To better understand how these variables are determined, consider a dataset
that contains the \textsc{Height}, \textsc{Weight}, \textsc{Age} and
\textsc{Score} values for a set of people. Using the back projection
interaction, we would like to experiment with the projection of an individual
with the attribute values $\textsc{Height}=174\text{, }\textsc{Weight}=68,
\textsc{Age}=30$, and $\textsc{Score}=8.5$. Suppose we constrain \textsc{Age}
to stay fixed (an equality constraint), \textsc{Score} to be between 8 and 10,
\textsc{Height} and \textsc{Weight} to be non-negative (inequality constraints)
using the Praxis interface.  Praxis would set $lb = [-174,-68,-\infty,-0.5]$
and $ub = [+\infty,+\infty,+\infty,1.5]$ for the inequality constraints.  For
the equality constraint on \textsc{Age}, Praxis sets $d=[0,0,30,0]$ and $C$ to
be a 4$\times$4 matrix with $[0,0,1,0]$ in its third row and zeros elsewhere.  

We now discuss a data exploration facilitated by our visual interactions in which
the underlying dimensionality reduction model is PCA. Drawing on earlier
work~\cite{Stahnke_2016}, we use the OECD Better Life dataset that contains eight
numerical socioeconomic development indices of 34 OECD member countries.  

\subsubsection{Example: OECD Better Life Index}
Zeynep is a data scientist working for a nonprofit organization focusing on
economic development. She wants to use the dataset to understand the
current situation of various countries in the world and validate her own
hypotheses. After importing the dataset into Praxis and choosing PCA as
the dimensionality-reduction method, Zeynep observes that the projection plane contains
three clearly separated clusters: (1) a large set of westernized
(mostly European) countries, (2) Portugal, Turkey, Mexico and Chile, and (3)
Korea and Japan. Noticing that Portugal is relatively distant from all other
European countries, Zeynep wants to understand which development indices
determine its position (Figure~\ref{fig:oecd}a). She selects the data point and
observes how, of the eight generated prolines, only four of them are long
enough to be visible---and they are associated (from the longest to the
shortest) to the features \textsc{StudentSkills},
\textsc{EducationalAttainment}, \textsc{SelfReportedHealth} and
\textsc{WorkingLongHours}. Immediately upon looking at the prolines, Zeynep
understands that the remaining four development indices have almost
no influence on  the current projection, while \textsc{StudentSkills} (the
longest proline) appears to be the most relevant feature. To verify this, she
tries to modify the feature value of \textsc{LifeSatisfaction} and observes
that, no matter how large the change, the data point associated to Portugal does
not move in the projection plane. On the other hand, slightly changing the
value of \textsc{StudentSkills} moves the point quickly along the
associated proline. By observing the direction of each proline, Zeynep
understands that features causing Portugal to be distant from the European
cluster are \textsc{EducationalAttainment} and \textsc{SelfReportedHealth},
while \textsc{StudentSkills} seems to be the main feature differentiating
it from Korea and Japan. Zeynep now wants to verify if a reasonable
increase in \textsc{StudentSkills} would make Portugal more similar to Korea.
While observing the feature distribution information on the associated proline,
she sees that Portugal would have to increase \textsc{StudentSkills} well beyond
the maximum value of the distribution.

Zeynep then focuses on another outlier country that has strong historical ties
to Europe but has never been part of it: Turkey (Figure~\ref{fig:oecd}b). She
selects the data point associated to Turkey and drags it towards one of the
closest European countries, Italy. Feature values of Turkey are updated through
(unconstrained) backward projection (Figure~\ref{fig:oecd}b, first table) and
Zeynep realizes that the country would have to increase almost all its
development indices to become more similar to Italy; only the
\textsc{WorkingLongHours} would have to decrease.  While dragging the data
point, Zeynep observes from the color of the highlighted prolines that
\textsc{StudentSkills}, \textsc{SelfReportedHealth} and
\textsc{EducationalAttainment} are positively intercorrelated (green color),
while \textsc{WorkingLongHours} is negatively correlated (red). Zeynep, wanting
to create a more realistic scenario, now assumes the Turkish government cannot
directly control indices such as \textsc{LifeExpectancy},
\textsc{SelfReportedHealth} and \textsc{LifeSatisfaction}, and sets an equality
constraint for these features.  However, the government can invest a certain
amount of money in education, with the plan of increasing the
\textsc{StudentSkills} index to 490 over the next five years. Zeynep sets the
inequality constraint on \textsc{StudentSkills} through a dedicated user
interface (Figure~\ref{fig:oecd}b, second table). She directly observes from
the feasibility map how the region of the projection plane around Italy is
reachable given the specified constraints.  Then Zeynep again moves Turkey
towards Italy through backward projection and observes how this time its
feature values  are updated to respect the user-defined constraints
(Figure~\ref{fig:oecd}b, second table).  While dragging the point, Zeynep
further validates her hypothesis by checking the changing position of
projection marks that indicate the current value of each feature with respect
to the distribution information encoded on prolines.

\begin{figure}[hb]
\includegraphics[width=\linewidth]{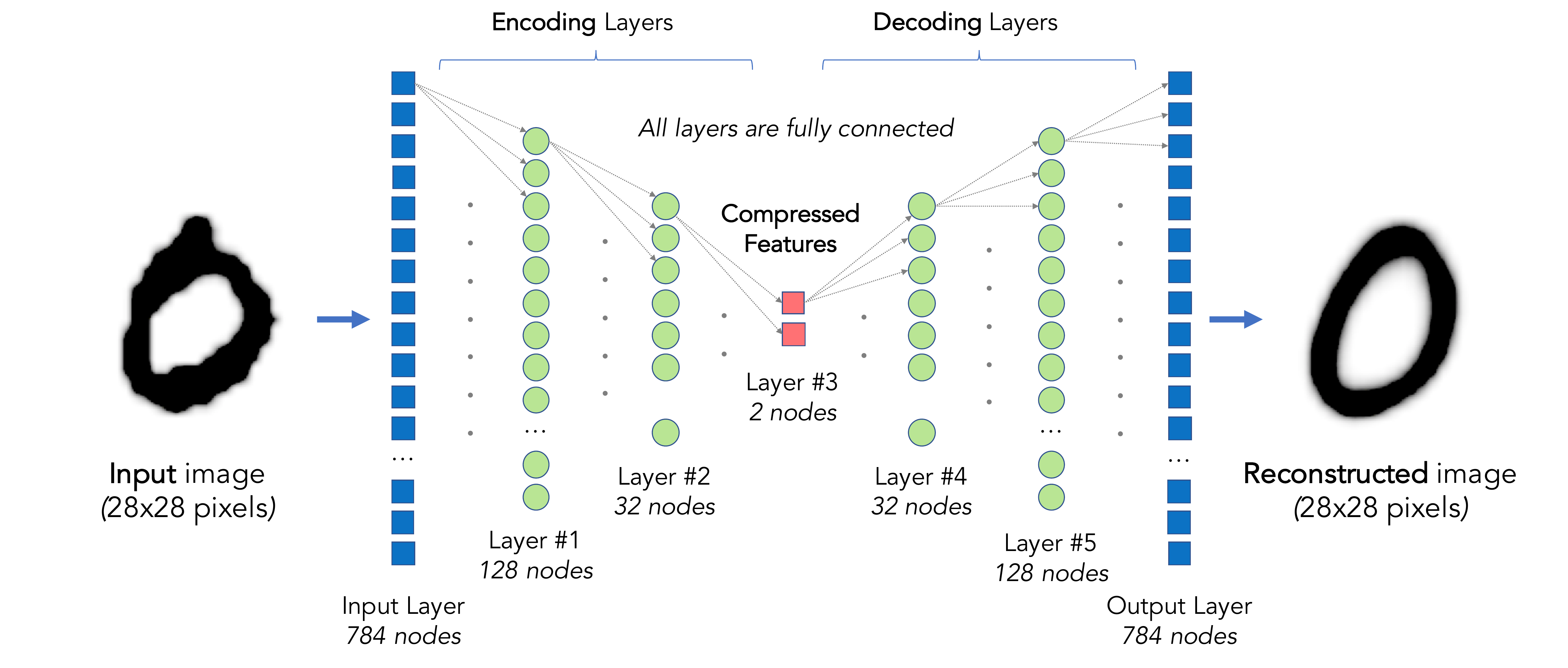}
\caption{\normalfont Architecture of the autoencoder trained on the example
  datasets. The first three hidden layers of the network represent the
  encoding function, which generates a compressed version of the input image.
  The last two hidden layers and the output layer represent the decoding
  function, which aims at reconstructing the original image from its 
  compressed representation. 
\label{fig:aemodel}} 
\end{figure}

\begin{figure*}[tbh] \centering
  \includegraphics[width=1\linewidth]{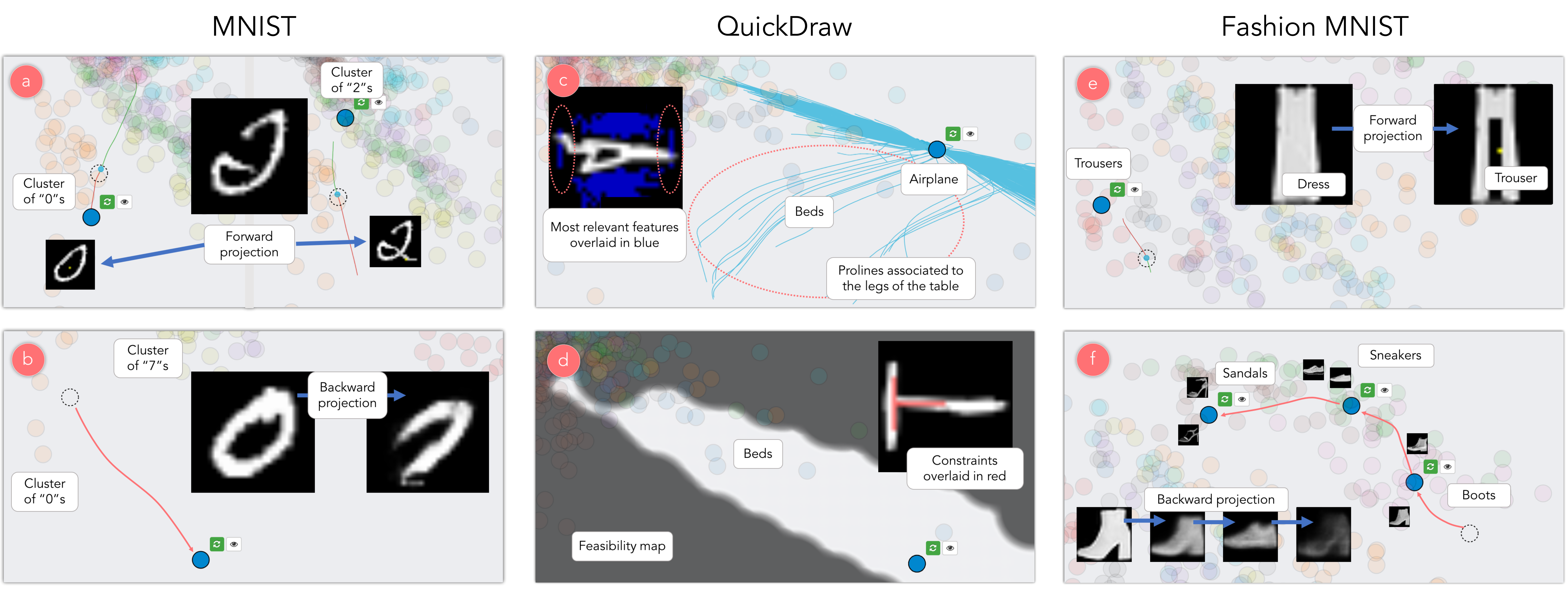}
  \caption{\normalfont Application of our framework to autoencoder-based 
    data exploration. MNIST: identification of relevant features for an unclear digit through
    prolines and forward projection (a); exploration of latent space at the
    borders of the projection plane (b). QuickDraw: prolines show that adding
    legs to an airplane would make it look more like a bed (c); by
    setting minimum constraints on the values of a set of pixels (colored in
    red), the feasibility map shows images similar to the current one (d).
    Fashion MNIST: the difference between dresses and trousers is explored through
    forward projection (e); key features differentiating various types of
    footwear are explored through backward projection (f).
    \label{fig:ae_examples}}
  \end{figure*}

\subsection{Application 2: Autoencoder}

In a second application, we demonstrate the framework interactions on
autoencoder-based DR. An autoencoder is an artificial neural network model that
can learn a low-dimensional representation (or encoding) of data in an
unsupervised fashion~\cite{rumelhart1986learning}. Autoencoders using 
multiple hidden layers with nonlinear activation functions can discover
nonlinear mappings between high-dimensional datasets and their 
low-dimensional representations. Unlike many other DR methods, an
autoencoder gives mappings in both directions between the data and
low-dimensional (latent) spaces~\cite{hinton2006reducing}, making it 
a natural candidate for application of the interactions introduced 
here. 

We compute forward projection by performing  an encoding pass on the trained
autoencoder for a user-modified input. To compute  backward projection, we
perform a decoding pass on the autoencoder for the user-changed output
projection. 

For the examples below, we trained an autoencoder model
(Figure~\ref{fig:aemodel}) with six layers of respective sizes  $(128,
32, 2, 32, 128, 784)$ from the first hidden layer to the output layer.  Our
examples below are from the dimensionality reduction of three image datasets: (1)
MNIST handwritten digit database, (2) Google QuickDraw (containing 50 million
drawings in 345 different categories), and (3) Fashion MNIST
(including images of clothing articles). All three datasets contain 28x28 pixel
grayscale images, represented as data vectors of 784 features.

\subsubsection{Example: MNIST}

Umberto, a data-science student,  wants to better understand an 
autoencoder-based DR of the well-known MNIST dataset. From Praxis, he observes
the scatter plot of dimensionally reduced data and notices that similar digits tend
to cluster together along different radial directions. He immediately notices an outlier (a ``2'') within a group of ``0''s and wants
to understand which key features (pixels) are determining its position in the
projection plane (Figure~\ref{fig:ae_examples}a).  Hovering on the image with
the mouse highlights the prolines associated with each pixels. Umberto observes
that lower pixel values at the center of the digit would move the data point
down, to within the ``0'' cluster; conversely, increasing pixel values in the
bottom-right corner of the image would move the data point up, towards the
other ``2''s. He hypothesizes that putting a tail on the digit would make it
look more like a ``2'' and verifies this through forward projection: coloring
the appropriate pixels (i.e., increasing their feature values) moves the data
point away from the cluster of ``0''s.

Umberto now observes how certain regions of the plane are quite dense, while
other regions contain very few data samples. He selects a data point from the
``0'' cluster and drags it into an empty region at the border of the projection
plane (Figure~\ref{fig:ae_examples}b). While dragging the point, he observes
how the pixel values update, showing alternately features of a ``0'' and of a
``7'', the two digits whose clusters are closest to this region of the plane.

\subsubsection{Example: QuickDraw}

Alice, a data scientist, is excited about the recent release of Google Quick
Draw dataset and would like to apply DR to a subset (10) of the image
categories in the dataset.  While performing her analysis with Praxis, she
decides to focus on an isolated data point corresponding to the image of an
airplane. By filtering its prolines in order to show only the 100 longest ones,
Alice can observe which features (pixels) are most influential in determining
the position of the image in the projection plane
(Figure~\ref{fig:ae_examples}c). In particular, she notices a smaller set of
prolines associated to two vertical strips of pixels, suggesting that an
increase in their values would move the airplane towards the cluster of
beds. Alice decides to perform forward projection and draws two ``legs'' at the
extremes of the airplane: she verifies that the data point has now moved in the
``bed'' cluster.

Continuing her search for outliers, Alice notices a bed that is apart from the others,
probably because it was drawn with only one leg. She wants to
see which other images in the dataset may contain a similar shape and uses the
brush to define inequality constraints on a set of pixels
(Figure~\ref{fig:ae_examples}d): all solutions with pixel values of
interest below a specific threshold are not considered acceptable.This way,
Alice can directly observe from the admissible region of feasibility map that 
this particular shape can be found only among bed images.

\subsubsection{Example: Fashion MNIST}
Mark, an analyst working for a growing apparel company, wants to
understand how fashion articles can be better categorized on the company's
website on the basis of  their image. He uses Praxis to apply autoencoder-based
dimensionality reduction on the dataset and notices that the projection plane
shows a clear separation of footwear data samples from clothes. The two
clusters are set apart by a diagonal group of bags whose images have a distinctive rectangular shape.

Mark wants to explore first the region of the plane containing shoes. He
selects a \textit{boot} data sample that appears as an outlier and brings it
towards the cluster of points with the same label through backward projection
(Figure~\ref{fig:ae_examples}e). Mark notices that the neck of the
boot becomes shorter, meaning that the boot's original height was above 
average. He hypothesizes that the height of a shoe is a critical factor in
determining its position in the projection plane. He then drags the same data
point close to the cluster of sneakers and watches the pixels of the image
modify so that the boot gets even shorter, validating his hypothesis. Moreover,
he notices that the boot has become flat, losing its characteristic heel. Finally, wanting 
to understand what distinguishes sandals from shoes and boots, he  continues dragging
the data point through backward projection. As he does so, he notices that the pixel
density of the image decreases significantly while moving along the main
diagonal of the projection plane. Indeed, sandals prove to be the class of
images with the fewest colored pixels, because  they are open shoes and require
less material for construction.

By observing the projection plane, Mark notes that dresses interestingly fall
in a region very close to the cluster of trousers. He then selects a dress and
observes that many of its (red) prolines are directed towards that cluster,
indicating a decrease in pixel values. By hovering these prolines, Mark
observes that they correspond to the central region of the dress image. Through
forward projection, he erases those pixels (i.e., sets their feature value to
zero), making the dress image look like a pair of trousers
(Figure~\ref{fig:ae_examples}e). He then observes how the data point quickly
moved into the cluster of trousers.  Finally, Mark asks himself where a skirt
would be positioned in the projection plane if the dataset contained one. So he
selects another dress data sample and erases the upper part of its image,
shaping a skirt. Noticing how little the data point has moved, he hypothesizes
that skirts would cluster with dresses.

\begin{figure*}[tbh] \centering
   \includegraphics[width=\linewidth]{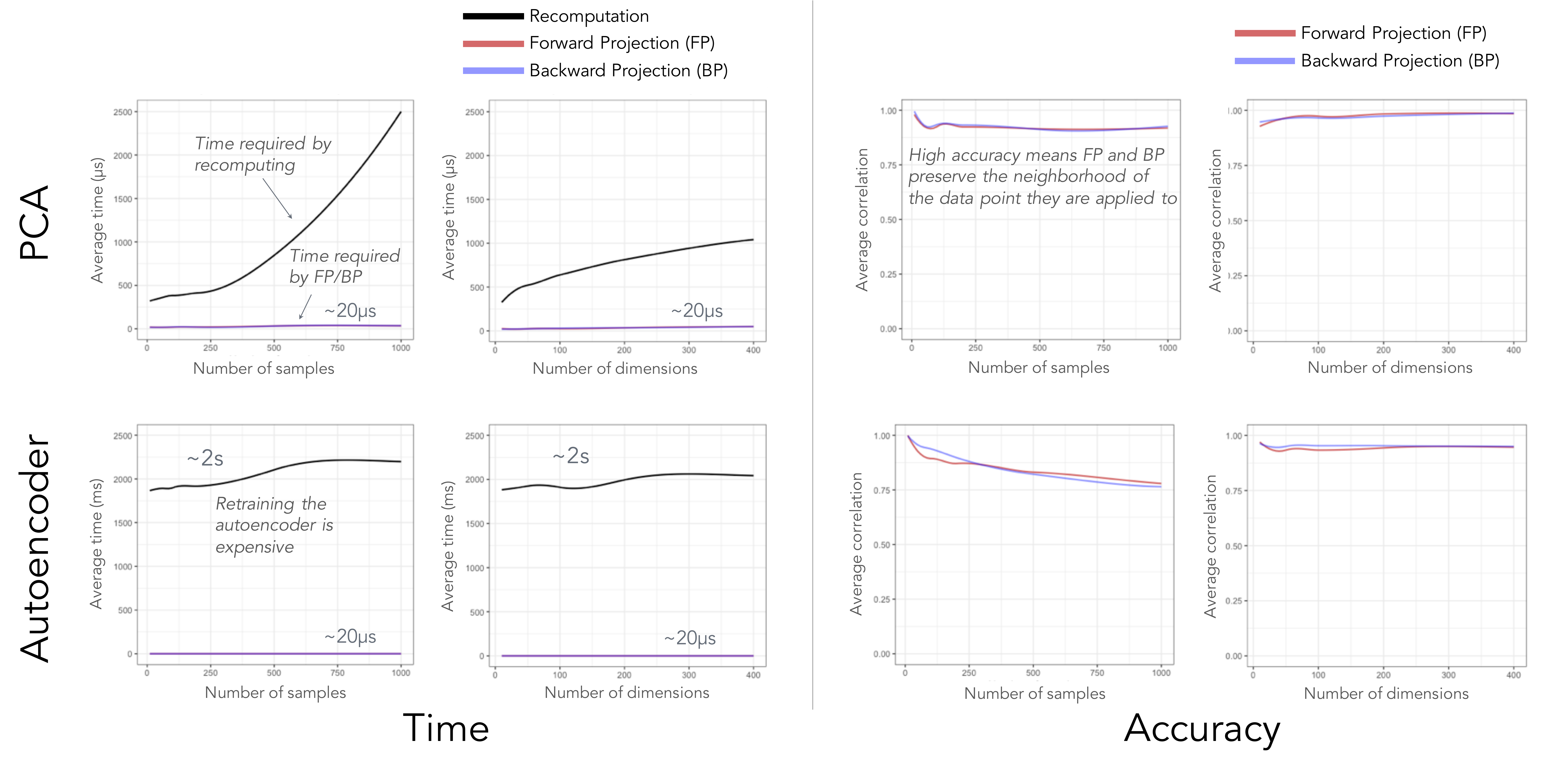}
   \caption{\normalfont Time and accuracy performance of forward and backward
     projections for PCA and autoencoder-based dimensionality reductions. Time
     performance results (left) show how out-of-sample extension outperforms
     recomputation, guaranteeing low latency even with increasing input size.
     Accuracy performance results (right) demonstrate how forward and backward
     projection provide a desirable tradeoff of accuracy and speed. 
     \label{fig:scalability}}
\end{figure*}

\section{Discussion}

\subsection{Scalability and Accuracy}

Scalability is a fundamental aspect of any visual interaction technique that
can handle large datasets. We design the visual interactions supporting our
framework with scalability in mind.  Our analysis indicates
(Figure~\ref{fig:scalability}) that our forward and backward projection
implementations for PCA and autoencoder provide a desirable tradeoff between
accuracy and speed.

To assess the performance of the framework interactions applied to PCA and
autoencoder dimensionality reductions (DRs), we measure the speed and accuracy
of forward and backward projections on varying number of data samples and
dimensions (see Supplementary Material for details and additional results). For
this, we first generate a synthetic dataset by sampling from a multivariate
Gaussian distribution with sufficiently large dimensions.  We then
programmatically induce a change of  $\sigma_i/8$ for forward projections
and  $m/80$ for backward projections, where $\sigma_i$  is the standard
deviation of the $i$th dimension in the dataset and $m$ is the width of the
projection plane. When varying the number of samples, we keep the number of
dimensions constant at 10. Conversely, when varying the number of dimensions,
we set the number of samples to 100.

Figure~\ref{fig:scalability} (left) suggests that the time required
to compute forward and backward projections for interactive applications
(\textasciitilde 20 microseconds) is not significantly affected by input size.
We note that recomputing the DR  without out-of-sample extension is several
orders of magnitude slower, especially in the case of autoencoder networks.

To measure accuracy, we compute two sets of neighbors on a data point for each
performance of our interactions: 1) the $k$-nearest neighbors in the projection
plane after performing a forward or a backward projection and 2) the
$k$-nearest neighbors in the projection plane after recomputing dimensionality
reduction on the multidimensional data. We set $k=10$ for the results shown in
Figure~\ref{fig:scalability}. Ideally, these two neighborhoods should contain
the same elements and the elements should have the same relative distance from
the data point on which the interaction is performed.  We use a correlation
index to quantify the similarity of these two neighborhoods, considering both
the number of overlapping elements and the rank order of their distances from
the point. Figure~\ref{fig:scalability} (right) shows  that the
accuracy decreases sublinearly with the input size. On the other hand, the
accuracy as measured by our correlation index is either unaffected or slightly
improved by the increasing number of dimensions, since the probability of a
single attribute change affecting the neighborhood formation and ranking
decreases with the increasing number of dimensions.

What about visual scalability? Rendering large number of prolines for
high-dimensional datasets can be slow and can clutter the projection view.
Sampling and bundling (aggregation)~\cite{da2016beyond} can help address this
problem. Taking a sampling approach, Praxis shows the top-k most ``important''
prolines when there are too many to draw (Figure~\ref{fig:ae_examples}c)).

An important consideration related to accuracy is \textit{trust}. While we
focus here on interpretability through dynamic reasoning and experimentation,
trust is also an important design criterion for model-based visual
analysis~\cite{Chuang_2012}. How best to convey the approximation accuracy of
our visual interactions to user without degrading computational and visual
scalability or inserting additional layer of complexity is an important avenue
of future research.

\subsection{Extending the Framework}

Although we focus on DR  here, our framework can apply to
black-box  models in general.  Neural network models with multiple hidden layers
(or deep neural networks, DNNs) are a particularly popular class of black-box machine
learning models, and they have recently achieved dramatic successes. However, formal
understanding of these models is limited, and the fact that they can
easily have millions  of parameters with increasingly intricate architectures only
exacerbates the problem.  Researchers have recently turned to
visualization to gain insights into deep  learning models. Prior work applying
visualization to  improve the understanding  and interpretability of deep
neural networks shows  patterns that fit well in our framework, attesting to
its  ecological validity and extensibility.

We can group DNN visualization approaches into two broad
classes~\cite{zintgraf2017visualizing}: The first approach is to visualize how
the network responds to a specific input in order to explain a particular
prediction by the network
(e.g.,\cite{bach2015pixel,li2015visualizing,mahendran2015inverting,shrikumar2016not,
simonyan2013deep,yosinski2015understanding,zeiler2014visualizing,
zhou2015cnnlocalization,
zintgraf2017visualizing}).
A common technique  used for convolutional neural networks for computer vision
applications is to \textit{occlude} parts of input images and \textit{observe}
how the output activation results (e.g., classification)
change~\cite{yosinski2015understanding,zeiler2014visualizing}. This is an
instance of a forward projection, changing the input and observing the output
change, albeit performed non-interactively in general, with one notable
exception~\cite{yosinski2015understanding}. Another common visualization in
this approach is saliency maps. Motivated similarly to  prolines, saliency maps
visualize which features (e.g., pixels) contribute to the output or any other
neural unit
activation~\cite{simonyan2013deep,zhou2015cnnlocalization,zintgraf2017visualizing}.

The second approach is to generate an input that maximally activates a given
unit or, say, class score to visualize what the network is looking for in
making predictions
(e.g.,\cite{erhan2009visualizing,
le2012building,lee2007sparse,
nguyen2016multifaceted,simonyan2013deep,
yosinski2015understanding}).
Techniques within this approach aim to synthesize input based on a maximization
constraint and can be considered an instance of backward projection. DNN
researchers also recognize  the importance of semantic constraints (e.g.,
natural image priors) in computing backward projections.

Visualization techniques within the two approaches above have been developed
primarily by machine-learning and computer-vision researchers to address their
research questions and to understand and communicate the behavior of their
models. These techniques are typically computed through  command- line
interaction and viewed as static images, with limited or no interactivity.
Applying dynamic interactions of our framework to DNNs can significantly
improve the effectiveness of the visualization techniques DNN researchers
already use. A full-fledged integration of backward projection interaction on
DNNs, one that interactively changes the activation  output of a neural unit
and observes the synthesized input, is challenging yet important future work.
Crucially, our framework can be useful for orienting future efforts in
supporting dynamic reasoning about DNNs. One class of models that would take
advantage of our interaction framework is generative and invertible
models~(e.g.,~\cite{chen2016infogan,
kingma2013auto,lample2017fader,perarnau2016invertible}).

Note that black-box models are essential abstractions representing the
modularization approach to problem solving.  Modularization is effective for
allocating human expertise and reducing monetary and cognitive costs, but
it decreases controllability and observability. Although contributions from the
broader HCI research community are necessary for improving the user experience
with machine-learning models, they are not sufficient.  Machine-learning models
need to be developed with built-in support (analogous to \textit{design for
debugging} or \textit{design for testability} in integrated circuit
design) for interpretability and explorability in mind.  Fortunately,
developing interpretable models is of a growing research interest
(e.g.,\cite{kim2015mind,
lei2016rationalizing,letham2015interpretable,ribeiro2016should}), but much still must be done in this direction,
particularly through close collaboration of HCI and machine-learning
researchers.

\section{Conclusion} 
We propose a new visual interaction framework that lets users dynamically
change the input and output of a dimensionality reduction  (DR) and observe the
effects of these changes. We achieve this framework through two new
interactions, forward projection and backward projection, along with two new
visualization techniques, prolines and feasibility map, that facilitate the
effective use of the interactions.  We apply our framework to principal
component analysis (PCA) and autoencoder-based DRs and give examples
demonstrating how our visual interactions can improve DR-based data
exploration. We show that the framework interactions applied to PCA and
autoencoders provide a desirable balance between speed and accuracy to sustain
interactivity, scaling gracefully with increasing data size and dimensionality.
Finally, we argue that our visual interaction framework can apply to black-box
machine learning models at large and discuss how our framework subsumes recent
approaches in visualizing deep neural network models.

Exploratory data analysis is an iterative process in which analysts essentially
run mental experiments on data, asking questions and (re)forming and evaluating
hypotheses. Tukey and Wilk~\cite{Tukey_1966} were among the first to observe
the similarities between data analysis and doing experiments. Of the eleven
similarities between the two that they listed, one in particular is relevant
here:``interaction, feedback, trial and error are all essential; convenience is
dramatically helpful.'' In fact, data can be severely underutilized (e.g.,
\textit{dead}~\cite{Haas_2011,Victor_2013}) without what-if analysis. However,
to perform data analysis as if we were running data experiments, dynamic visual
interactions that bidirectionally bind data and its visual representations must
be among our tools. Our work here is a contribution to performing visual
analysis in a way similar to running experiments.

\section{Acknowledgments} 
The authors thank Paulo Freire for inspiring the name ``Praxis.''
\bibliographystyle{SIGCHI-Reference-Format}
\balance
\bibliography{paper}
\end{document}